# Causal Estimation of Stay-at-Home Orders on SARS-CoV-2 Transmission

M. Keith Chen[1]*, Yilin Zhuo[1], Malena de la Fuente[1], Ryne Rohla[1], Elisa F. Long[1]

[1] UCLA Anderson School of Management, Los Angeles, CA 90095, USA.

*Correspondence to keith.chen@anderson.ucla.edu

This working paper was updated on 5/11/2020, the latest version is here.

**Abstract:** Accurately estimating the effectiveness of stay-at-home orders (SHOs) on reducing social contact and disease spread is crucial for mitigating pandemics. Leveraging individual-level location data for 10 million smartphones, we observe that by April 30th—when nine in ten Americans were under a SHO—daily movement had fallen 70% from pre-COVID levels. One-quarter of this decline is causally attributable to SHOs, with wide demographic differences in compliance, most notably by political affiliation. Likely Trump voters reduce movement by 9% following a local SHO, compared to a 21% reduction among their Clinton-voting neighbors, who face similar exposure risks and identical government orders. Linking social distancing behavior with an epidemic model, we estimate that reductions in movement have causally reduced SARS-CoV-2 transmission rates by 49%.

**One Sentence Summary:** Stay-at-home orders modestly reduce movement—by 16% with sizeable partisan differences—and decrease SARS-CoV-2 transmission by 7%.

To mitigate the severity of the ongoing COVID pandemic, governments have launched a range of social distancing policies including, by early April, mandatory stay-at-home orders (SHOs) in forty-five U.S. states and the District of Columbia. Additional gathering restrictions and closures of restaurants, schools, and non-essential businesses are gradually being lifted, with at least 18 states poised to ease some restrictions and partially reopen the local economy (*1*), despite official forecasts that daily COVID deaths could reach 3,000 by June 1st (*2*). Expert recommendations emphasize the need for first increasing test availability while strengthening essential worker training, before incrementally relaxing social distancing policies (*3*). As restrictions are lifted, identifying which policies most effectively reduce virus transmission can help lower the risk of a second wave of infections, and mitigate its spread should one arise (*4*). Utilizing device-level geolocation data for 10 million U.S. smartphones to measure individual movement, combined with precinct-level election outcomes and block-group level demographics, we estimate the causal effect of SHOs on daily movement—and what drives non-compliance. We combine these estimates with an epidemiological model, to estimate the net impact of SHOs and other social distancing mandates on SARS-CoV-2 transmission.

Early signs indicate that social distancing can substantially "flatten the curve", evidenced by case reductions following severe travel restrictions in China (*5, 6*). Lockdowns in several European countries are estimated to halve the effective reproduction number $R_e$, while banning public events, closing schools, and encouraging social distancing reduce transmission by 5-20% (*7*). Measures of SHO effectiveness in the United States vary substantially, ranging from an 8% reduction in mobility (*8*) to a 19% decrease in visits outside the home (*9*), with other studies estimating 35-45% fewer cases two weeks after a SHO is implemented (*10, 11*). In harder hit regions including Seattle (*12*), San Francisco and Los Angeles (*13*), and New York City (*14*),

SHOs are thought to cut transmission rates by 50% or more. Given an initial basic reproduction number $R_0$ of 2.5 to 3.0—and in the absence of widespread vaccination or herd immunity—transmission reductions of 60-67% are needed to fully curtail the pandemic (*15*).

Heterogeneity in adherence to social distancing mandates may stem from variations in local infection rates and healthcare capacity, differences in actual or perceived transmission risk, and misinformation or mistrust in government-issued orders, which may be exacerbated by selective media exposure (*9, 16*). Survey evidence indicates that partisan identity strongly correlates with attitudes towards COVID risk and acceptance of social distancing guidelines (*16, 17*). Republican governors issue distancing policies 1.7 days later, on average, than their Democrat counterparts, a delay that reaches 2.7 days in states with a large Trump electorate, even after controlling for local case counts (*18*). Google searches for COVID-related terms are lower in Trump-majority counties, although this difference attenuates following reports of self-quarantining by prominent Republicans (*19*). By late April, 38% of Republicans and 5% of Democrats report that "strict shelter-in-place measures are placing unnecessary burdens on people and the economy and are causing more harm than good" (*20*).

Partisan perceptions toward COVID appear to translate into actual behavior differences. Several studies document partisan gaps in social distancing using county-level geolocation data, including frequency of visits to points of interest (*21*), average daily distance traveled (*19*), or proportion of residents remaining entirely at home (*22*). Compliance with state-issued distancing orders further drops among residents in the opposite political party as their governor (*22*), an observation supported by survey data (*23*).

Accurately estimating the degree to which partisan identity or demographics modify social distancing behavior is a challenging empirical question. Current studies rely on geolocation data aggregated to the U.S. state (*10, 13*), county (*8, 21*), or populated regions in China (*5, 6*). Examining correlations between aggregate distancing behavior and political make-up can misestimate partisan responses to stay-at-home orders because Democrat-leaning counties account for the overwhelming majority of COVID cases, at all phases of the epidemic (Fig. S5). Analyzing movement behavior at the *individual-level* mitigates this potential confound, by restricting comparisons to Trump and Clinton voters who reside in nearby precincts within the same county— and thus face similar exposure risks, local mandates, and political leaders (Fig. S2).

We utilize smartphone device-level data to compute the total distance traveled outside the home each day, totaling more than 400 million person-day observations. By April 30th, average movement had declined 70% from pre-pandemic levels, with SHOs causally accounting for one-fourth of the decline, a 16% decrease ($1-e^{-0.170} = 0.16$, $p<0.0001$), as many residents severely curtailed movement outside the home before obtaining a formal SHO (Fig. 1). Nearly 75% of statewide SHOs occurred within two weeks of mandatory school, restaurant, and non-essential business closures (*1, 24*). The combined effect of these closures along with gathering restrictions and SHOs was a 31% reduction in travel ($p<0.0001$, Table S1, column 1), implying that less than half the reduction in movement is directly attributable to SHOs, closures and gathering restrictions. Consistent with other studies (*10, 13*), we find that residents of Democrat-leaning states respond more strongly to distancing orders (Fig. 2). New York and New Jersey, for instance, witness net reductions in movement of nearly 80%, perhaps unsurprisingly as these states account for 40% of all U.S. cases.

Our dataset includes detailed election results for approximately 172,000 voting precincts, the finest spatial aggregation publicly reported (*25*). Our difference-in-differences analysis indicates that, following a stay-at-home order, likely Clinton voters reduce daily movement by 21% (*p*<0.0001) whereas Trump voters reduce movement by 9% (*p*<0.0001) (Table S1, column 2). Comparing likely political affiliation to other key demographics in a multivariate regression reveals greater post-SHO movement declines among wealthier, older, urban, more educated, and non-white Americans (Table 1). Even after controlling for these covariates, the partisan wedge in movement reductions persists and remains statistically significant, with the difference between likely Trump and Clinton voters comparable to an increase in annual income of $50,000 or more. Mapping the changes in travel behavior before and after a SHO highlights the visual correlation between regions most afflicted with COVID and movement reductions (Figs. 3, S3-S4).

To test the robustness of our empirical specification to unobserved regional differences in factors like COVID prevalence or health resources, we compare SHO compliance among individuals living within the same Geohash-3 (a 150km×150km square grid), but on opposite sides of a county boundary. Although post-SHO movement reductions drop to 14% (Clinton voters) and 3% (Trump voters), the difference-in-differences remains similar in magnitude (Table S1, column 3). Comparing residents in the same county but different precincts indicates a 15 percentage-point wedge in movement reductions between likely Trump and Clinton voters (Table S1, column 4).

Our second set of regressions examines to what extent post-SHO reductions in movement modify local (within county) SARS-CoV-2 transmission rates, using a standard Susceptible-Exposed-Infectious-Removed (SEIR) epidemic model. A naïve regression that omits observed daily movement—the causal pathway between SHOs and infection transmission—would attribute a 52% reduction in transmission rates to SHOs (Table S2, column 1). Interpreting this correlation is complicated, as it conflates stay-at-home orders causally reducing cases, and an uptick in cases prompting states to enact orders. We aim to correct for this issue by directly estimating the channel through which the policy takes effect, a technique known as a "front-door" adjustment (*26*).

We estimate that, averaged across the course of the epidemic, each infected person transmits the virus to 1.2 other residents in their county, controlling for county differences and time trends in case counts (column 2). Here, we apply a 10-day infectiousness period following a 5-day asymptomatic period (*4, 15*). By regressing new COVID diagnoses on observed daily (lagged) movement, we estimate that for every 10% decrease in meters traveled by residents, the local transmission rate decreases by 4.3% (column 3). Combining this estimate with our earlier results produces a causal estimate of the reduction in local transmission due to all orders of 15%, with one-half of the reduction (7%) directly attributable to SHOs.

The discrepancy in transmission reductions between the naïve regression and our causal estimate is evident from the downward trend in individual movement, well before SHOs went into effect (Fig. 1). People reduced movement as the pandemic worsened, driven (presumably) by the perceived severity of the crisis, measurably filtered through a partisan lens. Applying these estimates to the evolving pandemic, we estimate an average within-county effective reproduction number of 1.14 among likely Clinton voters, and 1.32 among likely Trump voters, suggesting that a Trump voter who contracts coronavirus infects 16% more people than a comparable Clinton voter.

While other studies also examine partisan patterns in social distancing behavior, our study highlights the importance of measuring movement patterns and political variation at more granular levels, particularly when an emerging epidemic disproportionately afflicts urban, coastal cities. Of the ten counties hardest hit by COVID, all but Suffolk, New York were won by Hillary Clinton, with an average two-party vote share of 72% (Fig. S5). By directly measuring the channel through which social distancing orders affect behavior, we obtain causal estimates of the impact of such orders on disease transmission, a result of importance to policymakers as these orders are lifted.

**Acknowledgments:** The authors thank Veraset for providing the dataset, and Peter Rossi for feedback.

**Funding:** none.

**Author contributions:** Conceptualization, methodology, analysis (MKC, EFL), data pre-processing, writing, plots (MKC, YZ, MDLF, EFL), map generation (RR).

**Competing interests:** none.

**Data availability:** Smartphone location data are available from Veraset.

# References

1. S. Mervosh, J. C. Lee, L. Gamio, N. Popovich, "See Which States are Reopening and Which are Still Shut Down," *The New York Times* (May 1, 2020).

2. "The Trump administration projects about 3,000 daily deaths by early June," *The New York Times* (May 4, 2020).

3. Edmond J. Safra Center for Ethics, "Roadmap to Pandemic Resilience: Massive Scale Testing, Tracing, and Supported Isolation (TTSI) as the Path to Pandemic Resilience for a Free Society," (Harvard University, 2020).

4. S. M. Kissler, C. Tedijanto, E. Goldstein, Y. H. Grad, M. Lipsitch, Projecting the transmission dynamics of SARS-CoV-2 through the postpandemic period. *Science*, eabb5793 (2020).

5. H. Fang, L. Wang, Y. Yang, "Human Mobility Restrictions and the Spread of the Novel Coronavirus (2019-nCoV) in China " *Working Paper No. 26906* (National Bureau of Economic Research, 2020).

6.  J. S. Jia *et al.*, Population flow drives spatio-temporal distribution of COVID-19 in China. *Nature*, 1-11 (2020).

7.  S. Flaxman *et al.*, "Report 13: Estimating the number of infections and the impact of non-pharmaceutical interventions on COVID-19 in 11 European countries,"  (Imperial College COVID-19 Response Team, 2020).

8.  S. Engle, J. Stromme, A. Zhou, "Staying at Home: Mobility Effects of COVID-19," *Working Paper*  (2020).

9.  M. Andersen, "Early Evidence on Social Distancing in Response to COVID-19 in the United States,"  (http://dx.doi.org/10.2139/ssrn.3569368, 2020).

10. R. Abouk, B. Heydari, "The Immediate Effect of COVID-19 Policies on Social Distancing Behavior in the United States,"  (http://dx.doi.org/10.2139/ssrn.3571421, 2020).

11. J. H. Fowler, S. J. Hill, R. Levin, N. Obradovich, "The Effect of Stay-At-Home Orders on COVID-19 Infections in the United States,"  (https://arxiv.org/abs/2004.06098, 2020).

12. N. Thakkar, R. Burstein, H. Hu, P. Selvaraj, D. Klein, "Social Distancing and Mobility Reductions have Reduced COVID-19 Transmission in King County, WA,"  (Institute for Disease Modeling, 2020).

13. A. I. Friedson, D. McNichols, J. J. Sabia, D. Dave, "Did California's Shelter-in-Place Order Work? Early Coronavirus-Related Public Health Effects," *Working Paper No. 26992* (National Bureau of Economic Research, 2020).

14. J. E. Harris, "The Coronavirus Epidemic Curve is Already Flattening in New York City," *Working Paper No. 26917*  (National Bureau of Economic Research, 2020).

15. R. M. Anderson, H. Heesterbeek, D. Klinkenberg, T. D. Hollingsworth, How will country-based mitigation measures influence the course of the COVID-19 epidemic? *The Lancet* **395**, 931-934 (2020).

16. L. Bursztyn, A. Rao, C. Roth, D. Yanagizawa-Drott, "Misinformation During a Pandemic," *Working Paper No. 2020-44* (University of Chicago, Becker Friedman Institute for Economics 2020).

17. S. Kushner Gadarian, S. W. Goodman, T. B. Pepinsky, "Partisanship, Health Behavior, and Policy Attitudes in the Early Stages of the COVID-19 Pandemic," (http://dx.doi.org/10.2139/ssrn.3562796, 2020).

18. C. Adolph, K. Amano, B. Bang-Jensen, N. Fullman, J. Wilkerson, "Pandemic Politics: Timing State-Level Social Distancing Responses to COVID-19," (https://doi.org/10.1101/2020.03.30.20046326, 2020).

19. J. M. Barrios, Y. Hochberg, "Risk Perception Through the Lens of Politics in the Time of the COVID-19 Pandemic," *Working Paper No. 27008* (National Bureau of Economic Research, 2020).

20. A. Kirzinger, L. Hamel, C. Munana, A. Kearney, M. Brodie, "KFF Health Tracking Poll - Late April 2020: Coronavirus, Social Distancing, and Contact Tracing," (Kaiser Family Foundation, 2020).

21. H. Allcott *et al.*, "Polarization and Public Health: Partisan Differences in Social Distancing during the Coronavirus Pandemic," *Working Paper No. 26946* (National Bureau of Economic Research, 2020).

22. M. Painter, T. Qiu, "Political Beliefs affect Compliance with COVID-19 Social Distancing Orders," (https://dx.doi.org/10.2139/ssrn.3569098, 2020).

23. K. Cornelson, B. Biloucheva, "Political Polarization, Social Fragmentation, and Cooperation During a Pandemic," *Working Paper No. 663* (University of Toronto, Department of Economics, 2020).

24. Crowdsourced COVID-19 Intervention Data. ([https://socialdistancing.stanford.edu/](https://socialdistancing.stanford.edu/), Stanford University, 2020).

25. R. Rohla. ([http://rynerohla.com/index.html/election-maps/2016-presidential-general-election-maps/](http://rynerohla.com/index.html/election-maps/2016-presidential-general-election-maps/), 2018).

26. J. Pearl, Causal Diagrams for Empirical Research. *Biometrika* **82**, 669-688 (1995).

27. "Coronavirus (COVID-19) Data in the United States," *The New York Times* (May 1, 2020).

28. M. K. Chen, R. Rohla, The Effect of Partisanship and Political Advertising on Close Family Ties. *Science* **360**, 1020-1024 (2018).

29. A. N. Glynn, K. Kashin, Front-Door Difference-in-Differences Estimators. *American Journal of Political Science* **61**, 989-1002 (2017).

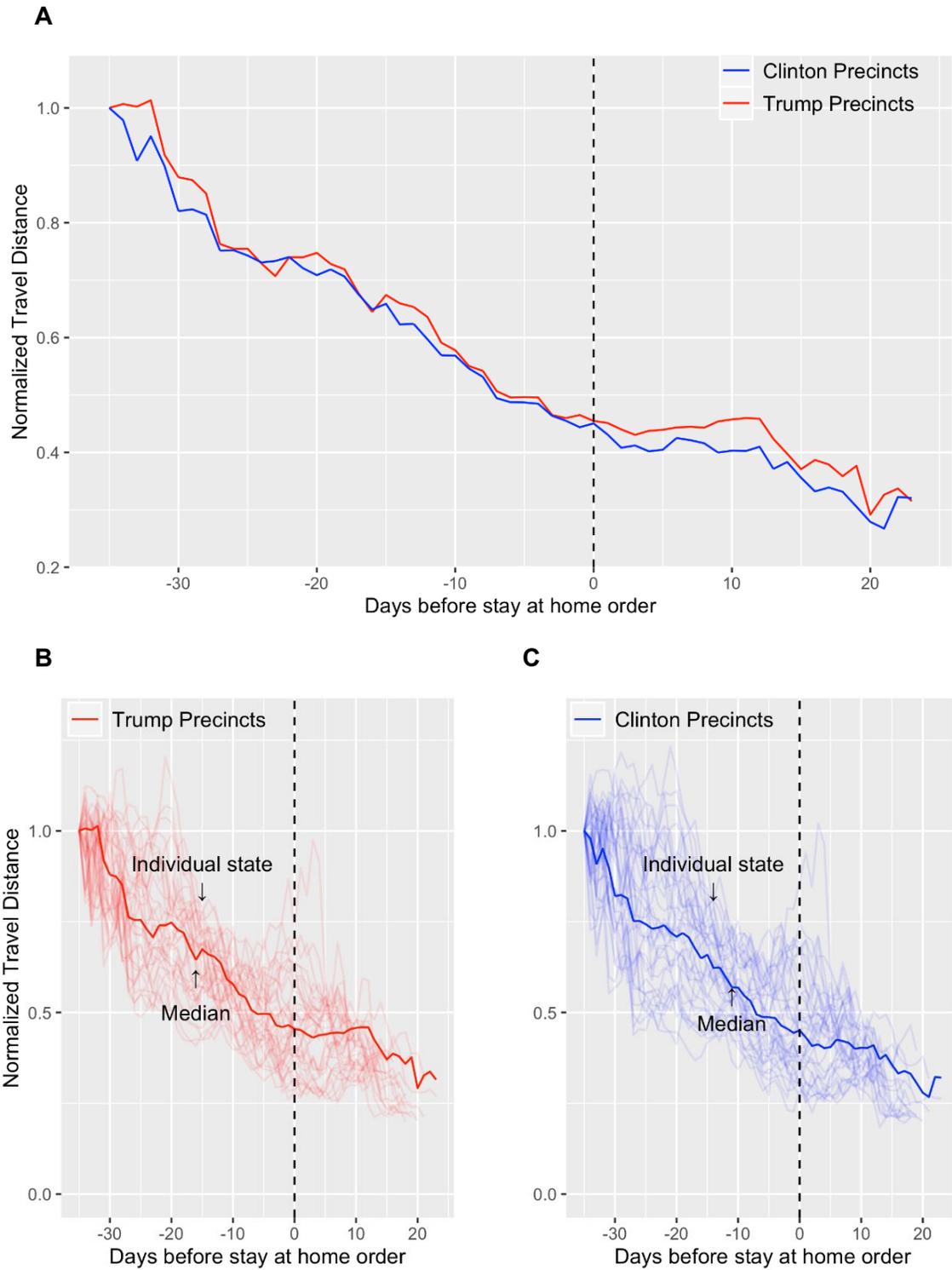

**Fig. 1.** Average daily travel distance before and after stay-at-home orders.

Average distance in (A) Trump- and Clinton-majority precincts based on 2016 election results and (B) by state, normalized to movement 35 days before each state's stay-at-home order was enacted.

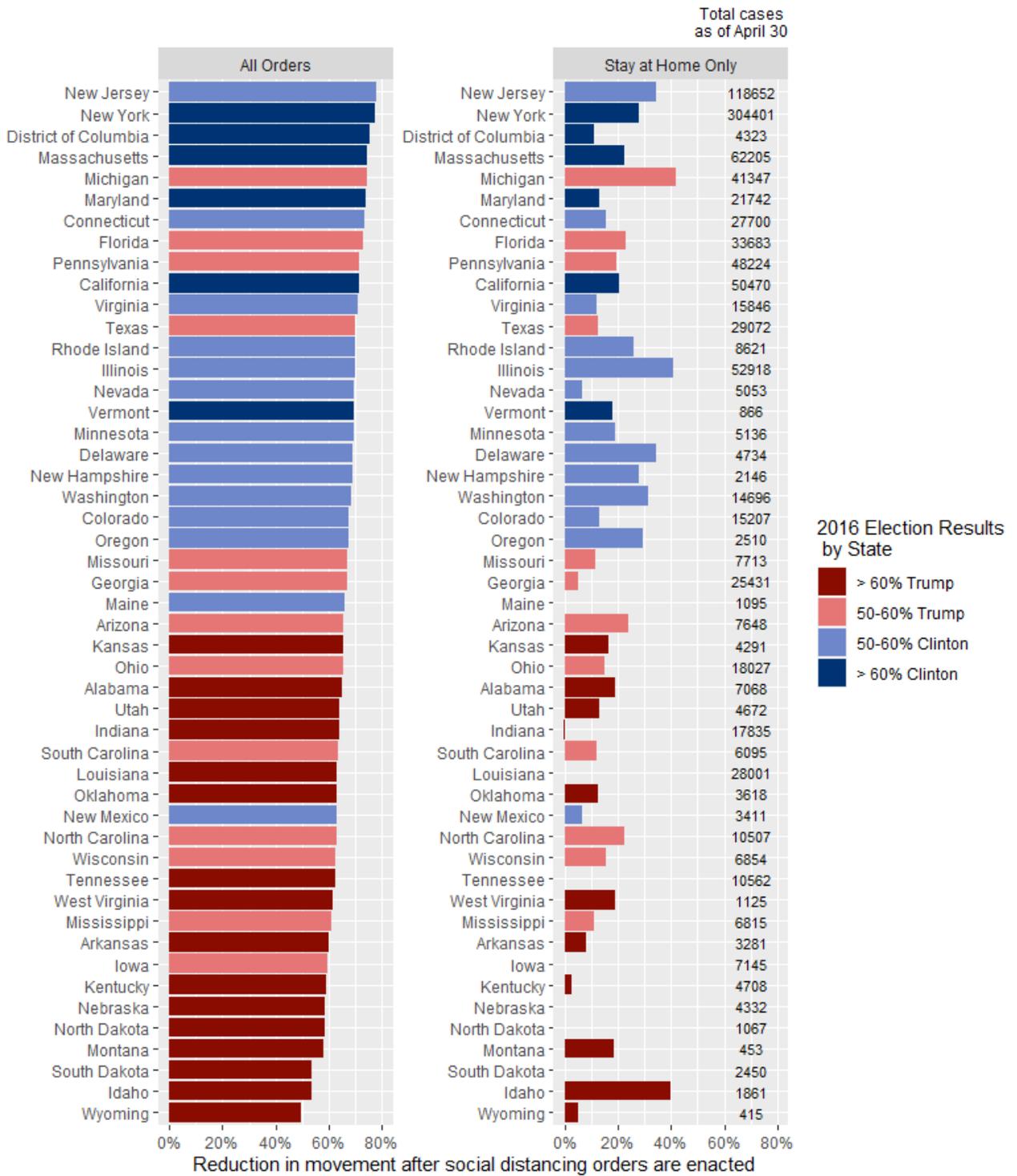

**Fig. 2.** State-level estimates of social distancing effectiveness.

Effectiveness estimates are derived from linear difference-in-difference regressions of stay-at-home and all orders (stay-at-home, gathering restrictions, and school, business and non-essential business closures). Total confirmed cases by state as of April 30 are given.

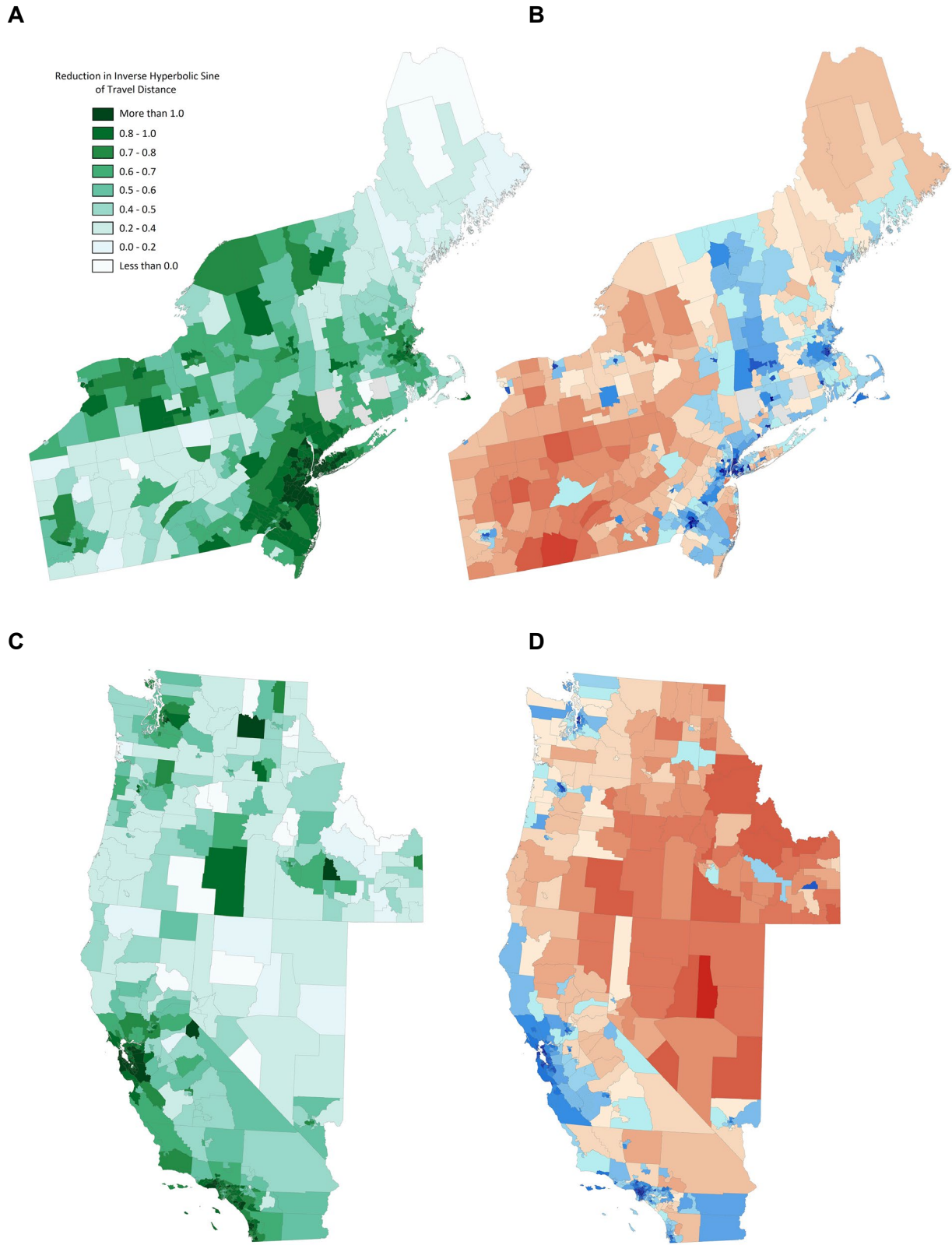

**Fig. 3.** Changes in movement following stay-at-home orders for COVID and 2016 Presidential election results in (A-B) New England and (C-D) West Coast, USA.

|  | Dependent variable: Daily Movement (arsinh meters) | | | |
|---|---|---|---|---|
| Independent variable | (1) | (2) | (3) | (4) |
| Stay-at-Home × Trump Share | 0.257*** | 0.181*** | 0.221*** | 0.122*** |
|  | (0.00189) | (0.00204) | (0.00217) | (0.00288) |
| Stay-at-Home × Pop. Density |  | -0.00367*** | -0.00427*** | -0.00362*** |
|  |  | (0.0000461) | (0.0000484) | (0.0000480) |
| Stay-at-Home × Age |  |  | -0.00105*** | -0.00207*** |
|  |  |  | (0.0000495) | (0.0000522) |
| Stay-at-Home × Income |  |  | -0.0270*** | -0.0238*** |
|  |  |  | (0.000178) | (0.000181) |
| Stay-at-Home × Bach. Deg. |  |  | -0.372*** | -0.368*** |
|  |  |  | (0.00522) | (0.00533) |
| Stay-at-Home × Grad. Deg. |  |  | -0.177*** | -0.215*** |
|  |  |  | (0.00616) | (0.00635) |
| Stay-at-Home × Black |  |  |  | -0.0369*** |
|  |  |  |  | (0.00297) |
| Stay-at-Home × Hispanic |  |  |  | -0.132*** |
|  |  |  |  | (0.00261) |
| Stay-at-Home × Asian |  |  |  | -0.498*** |
|  |  |  |  | (0.00567) |
| Controls for Govt. Restrictions | Yes | Yes | Yes | Yes |
| Observations | 423,803,900 | 423,631,796 | 416,267,558 | 416,267,558 |
| $R^2$ | 0.446 | 0.446 | 0.447 | 0.447 |
| Fixed Effects | Person, Date | Person, Date | Person, Date | Person, Date |

**Table 1.** Partisan and demographic predictors of stay-at-home compliance.

Each linear difference-in-differences regression estimates the effect of government stay-at-home orders on daily movements of smartphone users, and the effect of demographics on compliance. Trump Share is precinct-level two-party vote share won by Donald Trump in 2016. All other demographics are from the 2017 American Community Survey, reported at the Census block-group level. Population density is thousands of residents per square mile; Age is median household age in years; Income is median annual household earnings in units of $10,000. Education (no/some college, bachelor's degree, or graduate degree), and Race (White, Black, Hispanic, Asian) are proportions within a block-group. All regressions include controls for other government orders: large-gathering restrictions, and non-essential business, restaurant, and school closures. Smartphone-user fixed effects control for time-invariant demographics, and date fixed effects control for day-to-day changes in aggregate movement. Standard errors are clustered by smartphone-user and reported in parentheses with significance levels: * $p<0.01$, ** $p<0.001$, ***$p<0.0001$.

# Supplementary Materials for

## Causal Estimation of Stay-at-Home Orders on SARS-CoV-2 Transmission

M. Keith Chen, Yilin Zhuo, Malena de la Fuente, Ryne Rohla, Elisa F. Long

Correspondence to: keith.chen@anderson.ucla.edu

**This PDF file includes:**

Materials and Methods
Figs. S1 to S5
Tables S1 to S2

**Materials and Methods**

Data Summary

We utilize four primary datasets in our analyses: (1) Individual-level smartphone GPS location data, (2) 2016 U.S. precinct-level election results, (3) U.S. county-level COVID diagnoses, and (4) State- and county-level official policies in response to the COVID pandemic.

Our device-level smartphone GPS data is from Veraset, a company that aggregates anonymized GPS data from smartphone applications. After filtering for consistently observed smartphones, the dataset is comprised of over 10 million smartphones spanning the continental U.S. The data comprise these smartphones' "pings" which record that phone's geo-location and timestamp. Pings are logged at irregular time intervals whenever a participating smartphone application requests location information. The modal time interval between two pings for a device is roughly 10 minutes. Our sample covers the last two weeks of February through April 30th. To ensure our sample includes only users who we observe consistently and reliably over the study period, we exclude smartphone-day observations with less than 10 location pings, and users with fewer than 20 observations in the critical period of March and April.

Precinct-level voting data, collected through internet scraping and electoral authorities, records the vote share won by Donald Trump in the 2016 U.S. Presidential election for 172,098 precincts across 99.9% of counties nationally (*25*). We infer each smartphone device's "home" voting precinct using all pings between 10:00 p.m. and 6:00 a.m. over our sample period; this is then linked to that precinct's 2016 two-party vote share to proxy for a user's political affiliation. Our measure of movement (and non-compliance with stay-at-home orders) is a user's daily travel distance, which is the total distance between all consecutive pings during a day.

COVID data is from the New York Times, and includes daily cumulative confirmed diagnoses at the county-level (*27*). County-level intervention policy data are collected from (*1, 24*) and include the dates when each specific policy took effect. We categorize all policies into one of five groups: school closures, non-essential business closures, public gathering restrictions, restaurant restrictions, and stay-at-home orders.

Our approach for combining individual smartphone pings with election data is similar to (*28*), who estimate the effect of post-election partisanship on Thanksgiving durations. Here, we merge these data with epidemic case data and government-issued policies, allowing us to identify off variation in the timing and location of policy enactment. We arrive at causal estimates of the impact of policies, such as stay-at-home orders, on case counts by directly observing the exact causal pathway: changes in social distancing behavior.

Epidemic Model

To estimate how social distancing affects future COVID cases, we employ a Susceptible-Exposed-Infected-Recovered (SEIR) compartmental model for each U.S. county. In essence, this model captures the population-level dynamics of SARS-CoV-2 transmission, progression from asymptomatic to symptomatic infection, and recovery or death. Uninfected individuals who are susceptible to the disease (i.e., those with no immunity triggered by prior infection) can contract the virus, after which they remain exposed with latent (asymptomatic) infection for a period of time with average duration $1/\lambda$, which we assume to be 5 days. Exposed individuals then progress to an infectious state, where they remain until either death (with rate $\mu$) or recovery (with rate $\delta$) occurs.

Within each county, susceptible individuals can contract coronavirus from infected individuals with a transmission rate of $\beta$. Importantly, asymptomatic individuals with latent infection can also transmit the virus to susceptibles at a lower contact rate of $\epsilon\beta$, where $0 < \epsilon < 1$. The model is represented by a system of nonlinear differential equations for each county:

$$\frac{dS}{dt} = -\beta S \frac{I}{N} - \epsilon \beta S \frac{E}{N}$$

$$\frac{dE}{dt} = \beta S \frac{I}{N} + \epsilon \beta S \frac{E}{N} - \lambda E$$

$$\frac{dI}{dt} = \lambda E - (\delta + \mu) I$$

$$\frac{dR}{dt} = \delta I$$

The basic reproduction number $R_0$– the average number of secondary cases caused by an infected individual in a predominantly susceptible population–is computed using a linearization of the SEIR model at the disease-free equilibrium.

$$R_0 = \frac{\epsilon \beta}{\lambda} + \frac{\beta}{(\delta + \mu)}$$

Here, the first term corresponds to new cases arising from the exposed (latent) period, which is the product of the net transmission rate $\epsilon\beta$ and the duration $1/\lambda$. The second term corresponds to new cases arising from the infected period, which is again the product of the transmission rate $\beta$ and the average duration of infectivity $1/(\delta + \mu)$. Note, the transmission rate $\beta$ can be thought of as the number of social contacts per unit time $n$ multiplied by the probability of viral transmission per contact $p$.

During a widespread pandemic, the presence of social distancing and other interventions aim to reduce the effective reproduction number $R_e$ below the critical threshold of 1. Suppressing onward disease transmission is most directly achieved through reducing $\beta$ via either reducing $n$ (e.g., through stay-at-home orders, school closures, non-essential business closures, and gathering restrictions) or reducing $p$ (e.g., through use of facemasks, personal protective equipment for healthcare workers, and maintaining adequate physical distance between individuals).

As a direct result of social distancing mandates, or people altering their behavior to reduce exposure risk, the transmission rate $\beta$ is changing over time. Although we cannot directly observe this parameter, we use a locally linear regression to estimate how changes in daily movement—and thus $\beta$—affect the number of new COVID diagnoses. This is akin to linearizing the nonlinear SEIR model equations to compute the reproduction number $R_e$ near the disease-free equilibrium. By multiplying our estimated $\beta$ by the average duration of infectivity, we can obtain an approximation for the intra-county reproduction number, which likely underestimates the aggregate cross-county value.

Estimation Strategy and Specifications

Our regressions are designed to estimate the causal impact of government distancing policies on SARS-CoV-2 transmission in the presence of any of the unobserved confounding factors depicted in the directed acyclic graph (DAG) in Fig. S1. These include any time-invariant individual demographics $d_i$ that both directly affect measured movement and sort individuals into counties with different policies (e.g., education, income), county-level characteristics $d_c'$ correlated with aggregate movement and COVID levels (e.g., population density), any regional time trends $a_{rt}$ that effect both local movement and county distancing policies (e.g. spring break in costal FL), and any national time trends $a_t'$ that correlate with both movement and COVID levels (e.g. changes in national testing guidelines).

Most important, any county-time $Confounds_{ct}$ that correlate the timing of distancing policies with new cases would bias the observed relationship (in the naïve regression) between policies and disease suppression. For instance, an increase in neighboring counties' COVID cases could lead to both more local cases and prompt aggressive county-level distancing policies. By explicitly restricting our estimates to those attributable to changes in individual movement—the causal path between distancing policies and disease transmission—our approach mitigates this potential for bias. What could still bias our estimates would be a strong county-time confound that effects both movement and new cases in that county at that time, but is not reflected in that county's distancing policies. Even in the presence of such a confound though, our estimate would likely still be closer to the true effect that what a naïve regression would measure.

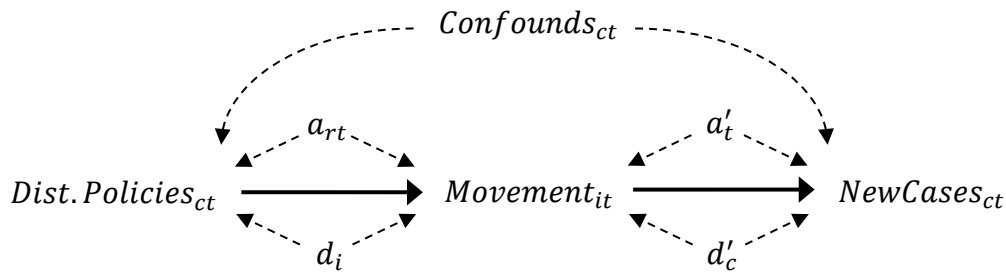

**Fig. S1.** Directed acyclic graph (DAG) modeling the causal effect of social distancing policies on new cases through the channel of reduced movement.

Regression 1 below estimates the effect of distancing policies on movement without conditioning on COVID cases. Regression 2 then estimates the effect of movement on local transmission (i.e., within an infected individual's county) controlling for any social distancing policies. Combining those estimates to arrive at the effectiveness of distancing policies at reducing SARS-CoV-2 transmission is equivalent to a "front-door" causal estimator as first developed by Pearl (*26*), and first used by Glynn and Kashin (*29*).

**Regression 1**
Our first regression estimates the degree to which policies (e.g., stay-at-home orders) reduce movement of individual residents.

$$Movement_{it} = \beta_0 + \beta_1 Policy_{it} + \beta_2 Policy_{it} \times TrumpShare_i + \alpha F_{irt} + \varepsilon_{it}$$

$Movement_{it}$ is the daily movement outside the home (measured in arsinh meters) traveled by individual $i$ on day $t$. $Policy_{it}$ is a vector of dummy variables for government distancing policies (stay-at-home orders, gathering restrictions, school closures, restaurant closures, and non-essential business closures) that individual $i$ experiences on day $t$. Note that most policies are enacted at the state-level, but a small fraction were originally implemented by individual counties (*24*). $F_{irt}$ are diff-in-diff-in-diff fixed effects for individual $i$ and spatial unit $r \times day$. We proxy individual $i$'s political affiliation using $TrumpShare_i$, the two-party vote share won by Donald Trump in individual $i$'s home precinct in the 2016 Presidential election. This regression estimates the causal impact of distancing policies on reducing resident movement, since not conditioning on COVID cases leaves $Movement_{it}$ and $Policy_{it}$ conditionally independent even in the presence of unobserved causes of distancing policies. Fig. S2 illustrates our identification strategy.

**Regression 2**
Our second regression estimates the local (within county) transmission rate of SARS-CoV-2, as derived in the SEIR model above.

$$NewCases_{ct} = \beta_0 + \beta_1 Infectious_{c\tau} \times Movement_{c\tau} + \beta_2 Policy_{c\tau} + \alpha F_{ct} + \varepsilon_{ct}$$

$NewCases_{ct}$ are new COVID diagnoses in county $c$ on day $t$. $Infectious_{c\tau}$ is the sum of previously diagnosed cases in county $c$ from time $\tau = t - (1/\lambda)$ and extending back the duration of infectivity $1/(\delta + \mu)$. $Movement_{c\tau}$ is the average daily movement of smartphone users in county $c$ on day $\tau$. $Policy_{c\tau}$ is a vector of dummy variables for government distancing policies in county $c$ on day $\tau$. $F_{ct}$ are difference-in-difference fixed-effects for county and day.

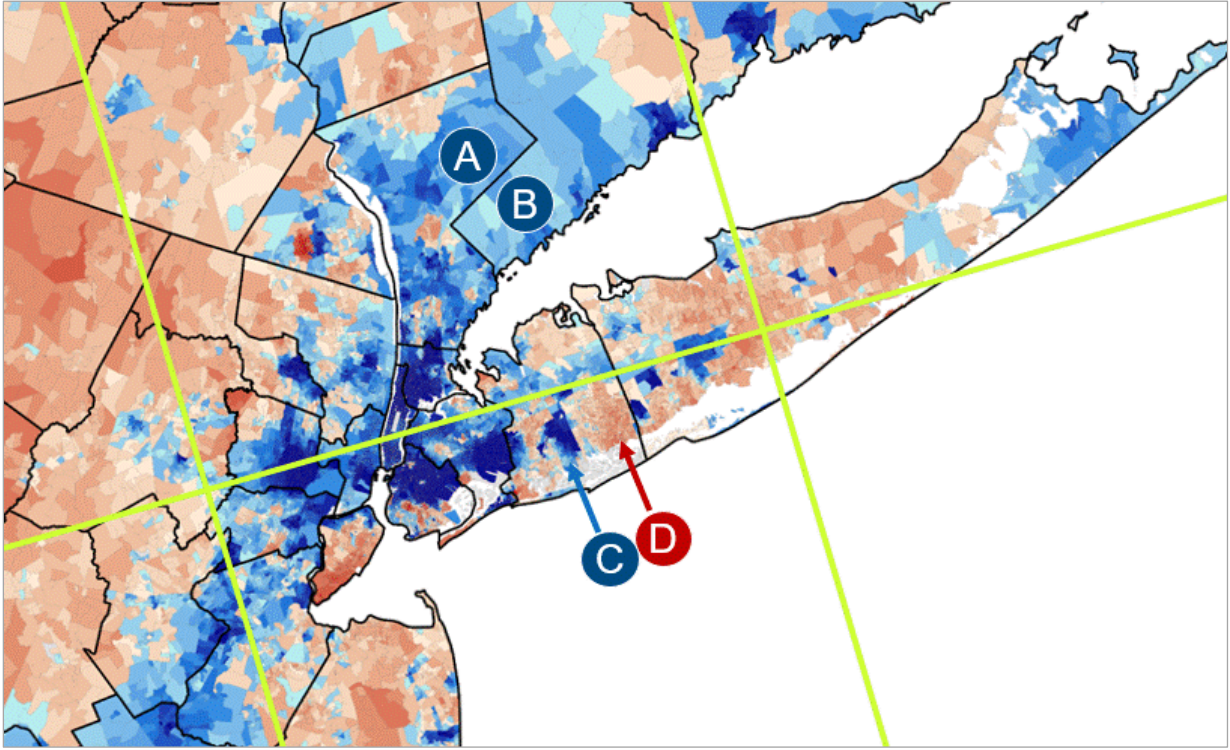

**Fig. S2.** Map of New York City metro area, with county boundaries outlined in black and geohash-3 boundaries in green. Letters indicate illustrative comparison groups from our regression specification. Points A and B represent individuals in different counties and states (Westchester, NY and Fairfield, CT), but with similar precinct Trump Shares. Points C and D represent individuals in the same county (Nassau, NY) but in precincts with differing Trump Shares. Controlling for geohash-3 regions further limits cross-county comparisons to those within close geographic proximity.

Map Generation

We merge our own precinct-level vote counts for the 2016 U.S. Presidential election, to construct the two-party vote share won by Donald Trump or Hillary Clinton (*25*). Using smartphone device-level geolocation data, we compute average daily distance traveled (arsinh meters) before and after stay-at-home orders are enacted, and aggregate these values to larger geographical units.

We plot the change in daily movement and two-party vote share, by county (in less densely populated areas) or PUMA (in more densely populated areas), to provide a visual correlation between the two measures. A PUMA, or Public Use Microdata Area, is a Census-designated sub-state grouping with at least 100,000 residents.

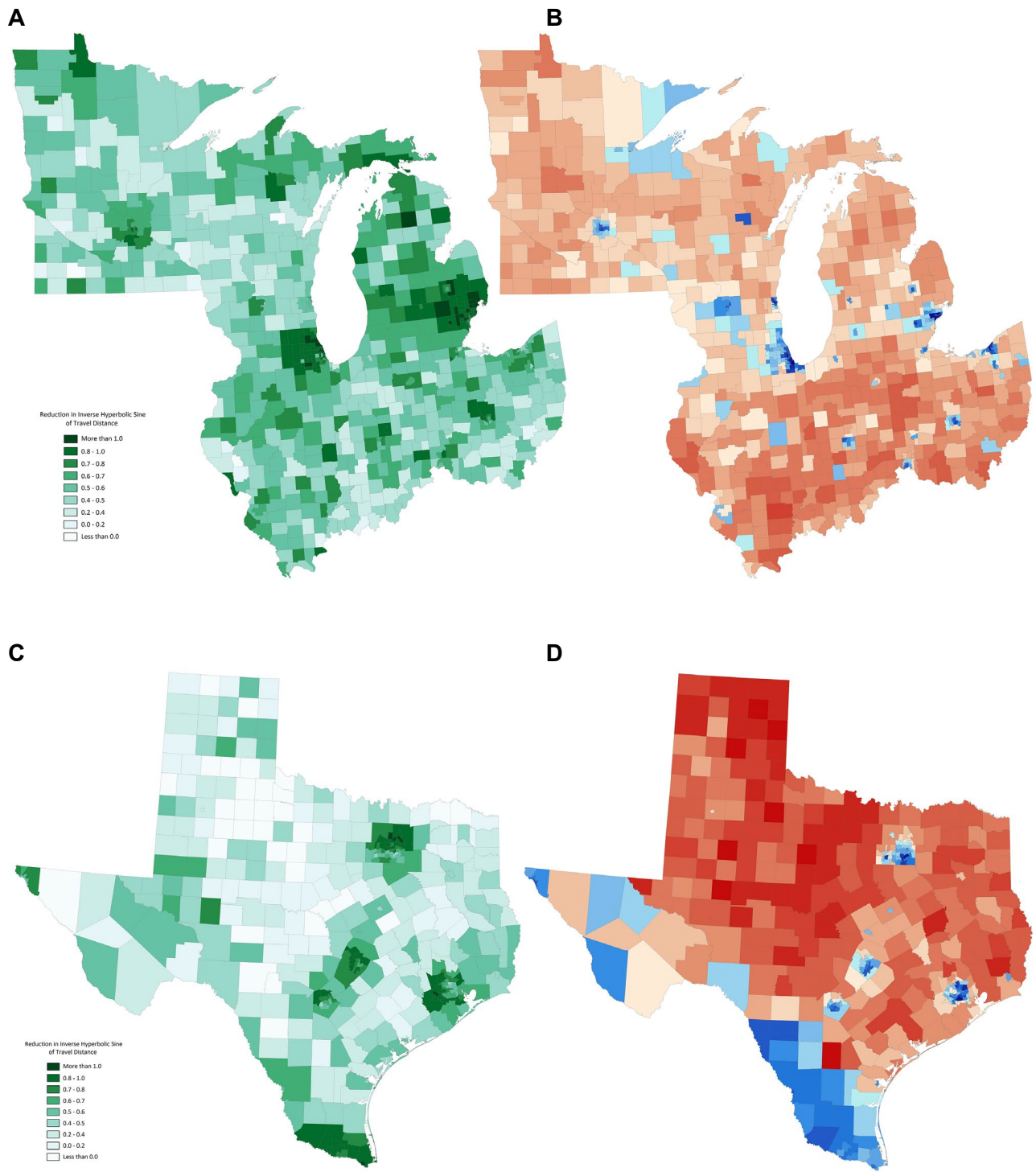

**Fig. S3.** Changes in movement following stay-at-home orders for COVID and 2016 Presidential election results in (A-B) Great Lakes region and (C-D) Texas, USA.

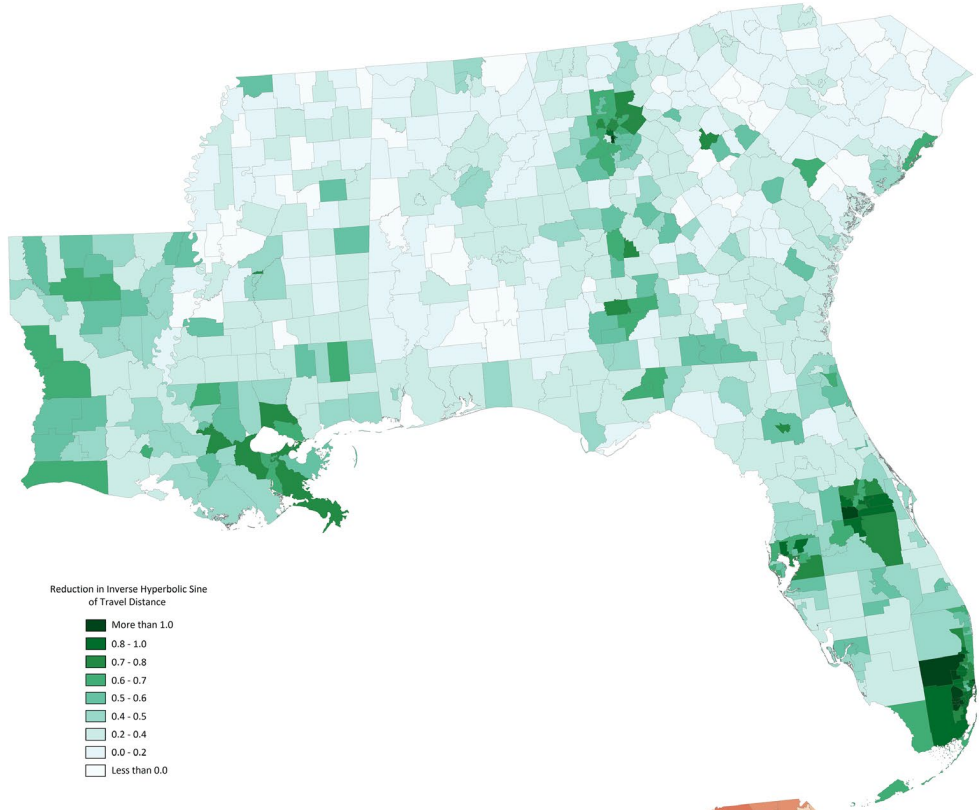
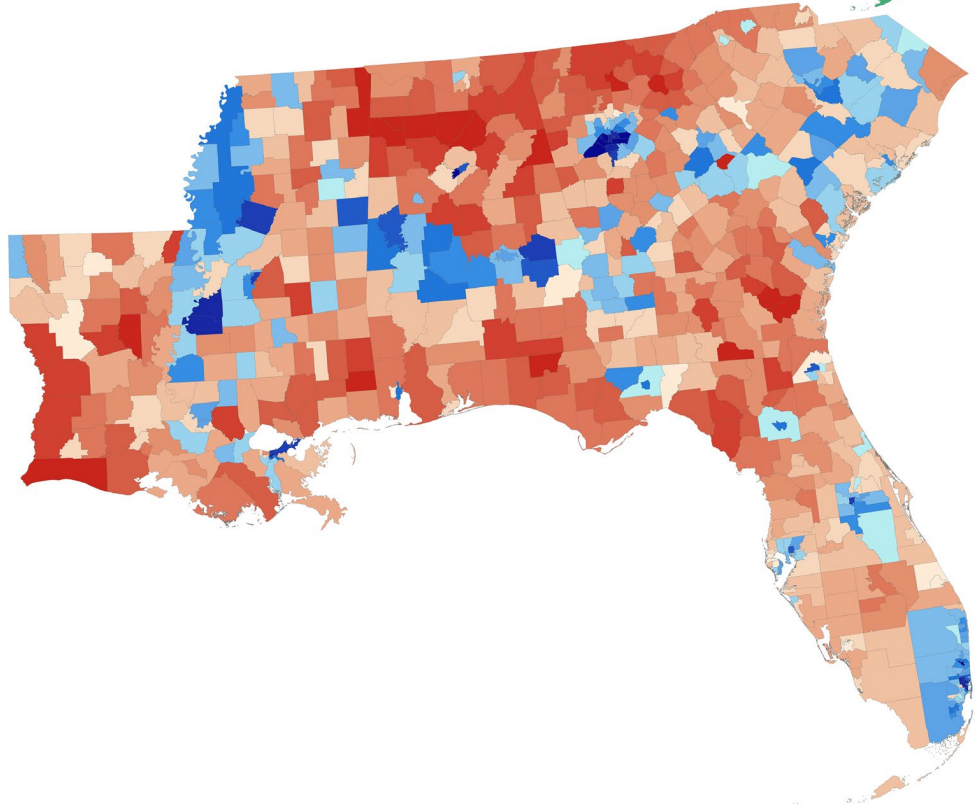

**Fig. S4.** Changes in movement following stay-at-home orders for COVID and 2016 Presidential election results in (A-B) Southern USA.

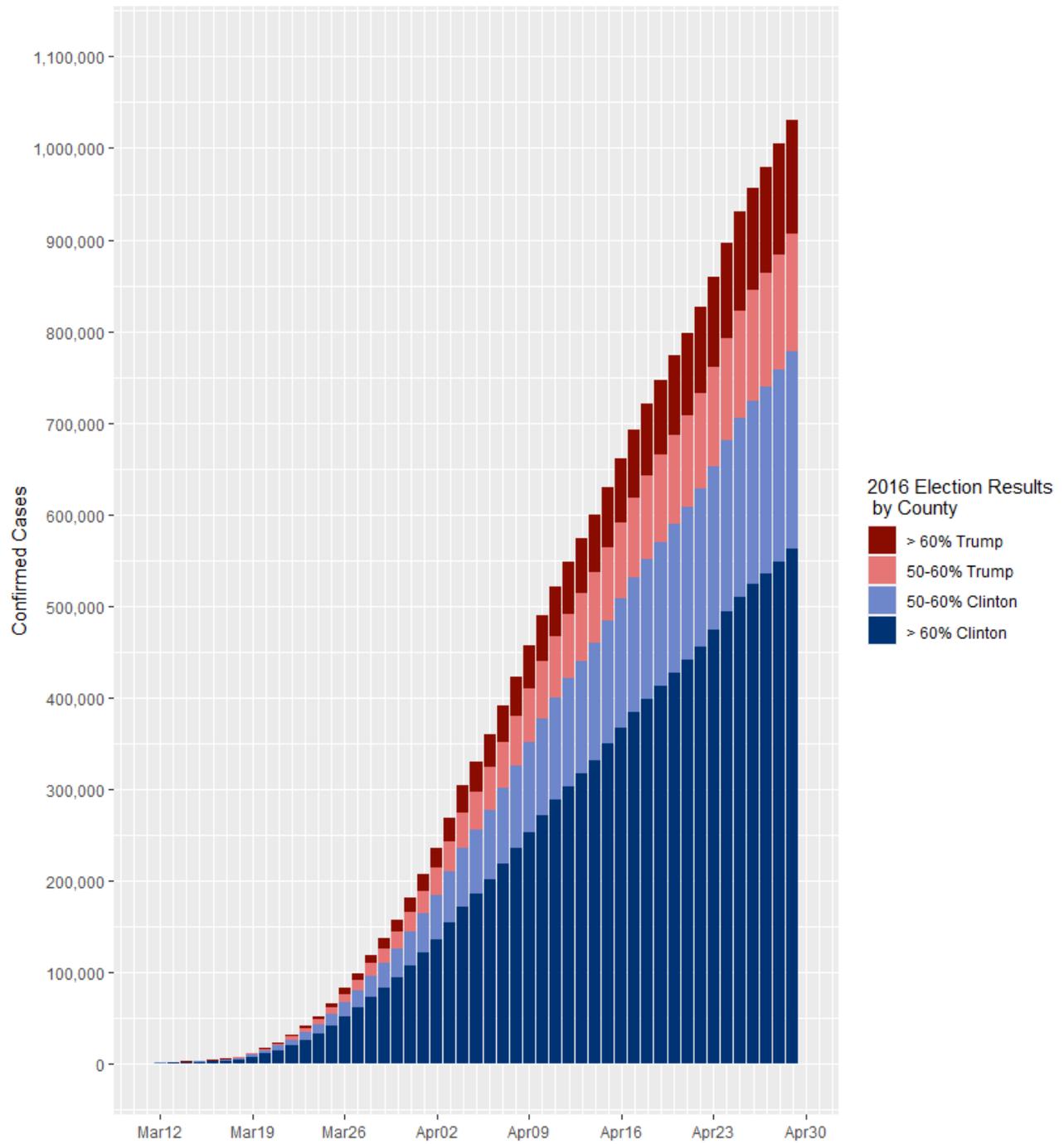

**Fig. S5.** Total confirmed COVID-19 cases among all U.S. counties as of April 30th.

|  | Dependent variable: Daily Movement (arsinh meters) | | | |
|---|---|---|---|---|
| Independent variable | (1) | (2) | (3) | (4) |
| Stay-at-Home Order | -0.170*** (0.000677) | -0.238*** (0.00161) | -0.156*** (0.00204) | NA |
| Gathering Restrictions | -0.0215*** (0.000606) | -0.0567*** (0.00144) | -0.00124 (0.00222) | NA |
| Non-Essent. Business Closures | -0.0990*** (0.000701) | -0.791*** (0.00177) | -0.0205*** (0.00206) | NA |
| Restaurant Closures | -0.0466*** (0.000772) | -0.0860*** (0.00182) | 0.00806* (0.00296) | NA |
| School Closures | -0.0290*** (0.000735) | -0.0441*** (0.00158) | 0.0139*** (0.00181) | NA |
| Stay-at-Home × Trump Share |  | 0.143*** (0.00263) | 0.129*** (0.00306) | 0.147*** (0.00419) |
| Gathering × Trump Share |  | 0.0596*** (0.00255) | -0.0246*** (0.00305) | -0.0818*** (0.00389) |
| Business × Trump Share |  | 0.00129 (0.00297) | -0.0728*** (0.00344) | -0.126*** (0.00469) |
| Restaurant × Trump Share |  | 0.0770*** (0.00307) | -0.00267 (0.00353) | -0.00177 (0.00453) |
| School × Trump Share |  | 0.0221*** (0.00272) | -0.0784*** (0.00309) | -0.219*** (0.00397) |
| Observations | 423,803,900 | 423,803,900 | 423,803,900 | 423,803,900 |
| $R^2$ | 0.446 | 0.446 | 0.448 | 0.450 |
| Fixed Effects | Person, Date | Person, Date | Person, Geo3 × Date | Person, County × Date |

**Table S1.** Effect of government orders on observed movement.

Each column is a linear difference-in-differences regression estimating the impact of government orders on observed daily movements of smartphone users, and the effect of partisanship on compliance (column 2-4). Trump Share is the precinct-level two-party vote share won by Donald Trump in 2016. All reported effect sizes are marginal net of all other orders. The direct effect of orders cannot be measured in column 4 because orders do not vary within counties on the same day; those coefficients are labeled NA. Smartphone-user fixed effects control for time-invariant demographics, and region × date fixed effects control for local day-to-day changes in aggregate movement. The sample includes 3,108 counties (or county equivalents) and 487 geohash-3 regions (a 150km × 150km square grid). Standard errors are clustered at the smartphone-user level and reported in parentheses with significance levels: * $p<0.01$, ** $p<0.001$, ***$p<0.0001$.

|  | Dependent variable: Daily New COVID Cases (county level) | | | |
| --- | --- | --- | --- | --- |
| Independent variable | (1) | (2) | (3) | (4) |
| Lagged Infectious Residents | 0.893 | 0.123*** | 0.0263 | 0.0287 |
|  | (0.637) | (0.00251) | (0.0560) | (0.0559) |
| Lagged Daily Movement |  |  |  | -3.212*** |
|  |  |  |  | (0.718) |
| Lagged Infectious Residents × Daily Movement |  |  | 0.0120 | 0.0117 |
|  |  |  | (0.00716) | (0.00715) |
| Lagged Infectious Residents × Stay-at-Home Order | -0.464*** |  |  |  |
|  | (0.0470) |  |  |  |
| Lagged Infectious Residents × Gathering Restrictions | -0.0677*** |  |  |  |
|  | (0.0147) |  |  |  |
| Lagged Infectious Residents × Non-Essent. Business Closures | -0.244 |  |  |  |
|  | (0.158) |  |  |  |
| Lagged Infectious Residents × Restaurant Closures | 0.174** |  |  |  |
|  | (0.0493) |  |  |  |
| Lagged Infectious Residents × School Closures | -0.163 |  |  |  |
|  | (0.495) |  |  |  |
| Controls for Govt. Restrictions | No | Yes | Yes | Yes |
| Observations | 155,400 | 155,423 | 155,400 | 155,400 |
| $R^2$ | 0.866 | 0.826 | 0.826 | 0.826 |
| Fixed Effects | County, Date | County, Date | County, Date | County, Date |

**Table S2.** SEIR epidemic model estimation.

Each column fits a SEIR model to daily county-level COVID cases. Independent variables are lagged for 5 days, representing the average asymptomatic period. Daily movement is measured in arsinh meters. Column 1 estimates a naïve regression examining how various government orders correlate with reduced transmission rates. Column 2 estimates the average local (within county) daily transmission rate per infectious individual. Columns 3-4 estimate the degree to which the local transmission rate depends on observed movement levels of county residents. Columns 2-4 control for the direct effect of government restrictions. In columns 3-4, a 10% reduction in daily movement is estimated to reduce transmission by 4-5% [ln(0.9) × 0.0120 / 0.0263) = -0.043]. F-tests for the joint significance of Lagged Infectious & Lagged Infectious × Movement are significant at the $p<0.0001$ level. The sample includes 3,108 counties (or county equivalents). Standard errors are clustered at the county-level and reported in parentheses with significance levels: * $p<0.01$, ** $p<0.001$, ***$p<0.0001$.